\documentclass[journal]{IEEEtran}
\usepackage{amsmath,amsfonts}
\usepackage{ragged2e}
\usepackage{algorithmic}
\usepackage{algorithm}
\usepackage{array}
\usepackage[caption=false,font=normalsize,labelfont=sf,textfont=sf]{subfig}
\usepackage{textcomp}
\usepackage{stfloats}
\usepackage{url}
\usepackage{verbatim}
\usepackage{graphicx}
\usepackage{cite}
\usepackage[utf8]{inputenc}
\usepackage{booktabs}
\begin{document}

\title{Cyberattack Detection in Virtualized Microgrids Using LightGBM and Knowledge-Distilled Classifiers}

\author{Osasumwen Cedric Ogiesoba-Eguakun,~\IEEEmembership{Member,~IEEE,} Suman Rath,~\IEEEmembership{Member,~IEEE}
        % <-this % stops a space
\thanks{Manuscript submitted: January 24, 2026. This work was conducted as part of the graduate research activities at the University of Tulsa.\\ 
(Corresponding author: Osasumwen Cedric Ogiesoba-Eguakun.)
 
O. C. Ogiesoba-Eguakun and Suman Rath are with the Department of Electrical and Computer Engineering, The University of Tulsa, Tulsa, OK 74104, USA (e-mail: oco1411@utulsa.edu, suman-rath@utulsa.edu).}

% Prof.  is with the Department of Electrical and Computer Engineering, The University of Tulsa, 74104, Tulsa, OK 74104, USA (e-mail: suman-rath@utulsa.edu
}% <-this % stops a space



\maketitle

\begin{abstract}
Modern microgrids depend on distributed sensing and communication interfaces, making them increasingly vulnerable to cyber–physical disturbances that threaten operational continuity and equipment safety. In this work, a complete virtual microgrid was designed and implemented in MATLAB/Simulink, integrating heterogeneous renewable sources and secondary-controller layers. A structured cyberattack framework was developed using MGLib to inject adversarial signals directly into the secondary control pathways. Multiple attack classes were emulated, including ramp, sinusoidal, additive, coordinated stealth, and denial-of-service behaviors.
The virtual environment was used to generate labeled datasets under both normal and attack conditions. The datasets trained Light Gradient Boosting Machine (LightGBM) models to perform two functions: detecting the presence of an intrusion (binary) and distinguishing among attack types (multiclass). The multiclass model attained 99.72\% accuracy and a 99.62\% F1-score, while the binary model attained 94.8\% accuracy and a 94.3\% F1-score. A knowledge-distillation step reduced the size of the multiclass model, allowing faster predictions with only a small drop in performance. Real-time tests showed a processing delay of about 54–67 ms per 1000 samples, demonstrating suitability for CPU-based edge deployment in microgrid controllers.
The results confirm that lightweight machine-learning–based intrusion-detection methods can provide fast, accurate, and efficient cyberattack detection without relying on complex deep-learning models. Key contributions include: (1) development of a complete MATLAB-based virtual microgrid, (2) structured attack injection at the control layer, (3) creation of multiclass labeled datasets, and (4) design of low-cost AI models suitable for practical microgrid cybersecurity. 
\end{abstract}


\section{Introduction}
\IEEEPARstart{P}{ower} systems are becoming more decentralized as renewable energy sources and distributed energy resources (DERs) continue to grow \cite{ogiesoba2023design}. Microgrids now play a key role in supporting flexible, resilient, and sustainable power delivery. However, the closer connection between physical equipment and cyber layers has expanded the attack surface, increasing exposure to cyber-physical threats that can disrupt stability, reduce data accuracy, and compromise operational safety \cite{srivastava2020microgrid}.

Microgrids depend on frequent measurement, communication, and control signals shared across supervisory and distributed controllers. These cyber connections create potential exposure to false data injection (FDI), denial of service (DoS), and stealth attacks \cite{liu2011false}. Such threats may disrupt control loops, lead to incorrect state estimation, or obstruct coordination \IEEEpubidadjcol among distributed generators. Therefore, reliable intrusion detection is necessary to maintain safe real-time operation.

Traditional intrusion detection methods that use rule-based logic or signal processing often fail to provide adequate protection, particularly when facing smart or hidden attacks. Recent research has therefore shifted toward data-driven cybersecurity solutions. Machine learning based detectors show strong potential because they can learn nonlinear patterns in power data and identify anomalies with higher accuracy \cite{beg2023review}. Progress in knowledge distillation (KD) \cite{hinton2015distilling} and gradient boosting models such as LightGBM \cite{ke2017lightgbm} support the development of fast and compact classifiers for edge-level protection.

Although deep learning models such as convolutional neural networks (CNNs) and long short-term memory networks (LSTMs) can achieve high detection accuracy, they often require large computational resources. This limits their use in real-time field controllers. In contrast, lightweight non-neural models such as gradient boosted decision trees can deliver similar accuracy with faster computation and lower memory demand. These strengths make them strong candidates for intrusion detection in microgrids with limited hardware.

This work develops a complete virtual microgrid in MATLAB Simulink using cyber-physical modeling methods described in \cite{srivastava2020microgrid}. The system includes diverse renewable generation sources and secondary control layers. A structured attack injection setup was created to introduce several adversarial behaviors, including ramp, sinusoidal, additive, coordinated stealth, and denial of service attacks, into the secondary controller interface. Labeled datasets from normal and attack conditions were then exported and used to train supervised LightGBM models for binary intrusion detection and multiclass attack classification. In contrast to earlier studies such as the PSERC S82G resiliency framework \cite{srivastava2020microgrid} and the cyber-secure distributed secondary control architecture of Rath et al. \cite{rath2020cyber}, which mainly describe the microgrid control structure and communication vulnerabilities, this work offers a complete cyber-physical simulation and a lightweight machine learning detection system. A high-resolution virtual microgrid model with a $2~\mu\text{s}$ switching step is built and includes controlled secondary control attack scenarios. Realistic labeled datasets are created across seven operating modes, and both binary and multiclass LightGBM models are trained and can run in real-time on a regular CPU. Knowledge distillation is applied to compress the model while keeping its accuracy. This makes it suitable for use on embedded microgrid controllers. To the best of our knowledge, no earlier study brings together high-fidelity cyber-physical modeling, structured secondary control attack paths, and efficient tree boosted models for real-time intrusion detection in microgrids.

The remainder of this paper is organized as follows: Section II focuses on the review of Cyberattack Detection Techniques for Microgrids. Section III describes the virtual microgrid and attack-injection architecture. Section IV describes the dataset generation and machine learning methodology. Section V reports the evaluation results. Section VI reviews computational efficiency, and Section VII concludes the paper.

\section{Related Work}
Research on microgrid cybersecurity has grown quickly because increased digital connectivity exposes control systems to malicious activity. Early studies showed that false data injection attacks can bypass traditional bad data detection, harming state estimation accuracy and threatening reliable operation \cite{liu2011false}. Other work demonstrated that denial of service attacks affecting communication channels can degrade secondary control performance, causing voltage and frequency instability in inverter-based networks \cite{mahmoud2015control}. These findings highlighted the need for stronger cyberattack detection in cyber-physical power systems.

Early defense strategies used statistical indicators, consistency checks, and signal processing techniques to identify unusual sensor behavior \cite{kundur2004definition}. These methods can handle sudden or large disturbances, but they struggle to detect coordinated or covert attacks. As a result, machine learning based intrusion detection systems have received considerable attention. For example, Sahoo and Kamwa used random forest models on real-time measurement data and reported stronger performance than conventional detection methods \cite{tuyizere2023machine}.

More recent studies have used deep learning to capture nonlinear and time-based patterns during cyber-attacks. Dey and Gupta developed a deep neural network framework that accurately detects malicious activity in smart grids \cite{qi2021detecting}. Gunduz and Das used a hybrid convolutional neural network (CNN) to detect malicious behavior in smart grid metering data, achieving high attack classification accuracy while keeping false alarms low \cite{gunduz2024smart}. Dehghani et al. developed a deep learning intrusion detection framework for Direct Current (DC) microgrids, showing strong ability to recognize cyber-attack signatures under changing operating conditions \cite{dehghani2021cyber}. Although these models provide strong accuracy, they require significant computational and memory resources, making real-time operation at edge controllers difficult.

To reduce complexity, lightweight learning models have been investigated. Gradient boosting methods such as LightGBM have shown strong results on structured grid data and can run well on standard Central Processing Unit (CPU) hardware \cite{ke2017lightgbm}. XGBoost has also been used to detect abnormal events in electric networks, giving good accuracy and low response time \cite{chen2016xgboost}. In addition, knowledge distillation has been used in cyber-physical settings to shrink large models, allowing smaller student models to learn from larger teacher models while using less computation \cite{hinton2015distilling}, \cite{gou2021knowledge}. Liu et al. showed that using knowledge distillation with federated learning (FL) can greatly reduce communication and computation demands, allowing real-time intelligence at edge cyber-physical system (CPS) nodes with limited resources \cite{liu2022efficient}.

Despite this progress, several limitations remain. Many current studies rely on static datasets instead of data generated from realistic microgrid simulations that include varied distributed energy resources. In addition, structured attack injection at the secondary control interface is often not implemented, limiting the evaluation of cyber-physical interactions. Finally, combining lightweight boosted models with knowledge distillation for multiclass cyberattack detection has received limited attention. These gaps motivate work that builds a MATLAB-based virtual microgrid, injects multiple types of cyberattacks at the control pathway, produces labeled datasets, and trains LightGBM classifiers enhanced through knowledge distillation for low-cost, real-time intrusion detection.

A summary of key methods is given in Table \ref{tab:taxonomy}. Although these techniques show good results with Machine Learning (ML) and Deep Learning (DL) models in different smart-grid settings, only a small number of studies have examined lightweight boosted models with knowledge distillation for multiclass cyberattack detection in virtual microgrid environments.

\begin{table}[!t]
\caption{Taxonomy of Selected Smart-Grid Cyberattack Detection Approaches}
\label{tab:taxonomy}
\centering
\begin{tabular}{|c|c|c|c|c|}
\hline
Ref. & Method & Target Area & Domain & Platform \\
\hline
[8]  & ML        & \begin{tabular}{@{}c@{}}Disturbance \&\\attack detection\end{tabular}   & \begin{tabular}{@{}c@{}}Power\\Systems\end{tabular}      & Python \\
\hline
[9]  & SSL-DL        & \begin{tabular}{@{}c@{}}Cyberattack \&\\detection\end{tabular}   & Smart Grid      & Python \\
\hline
[10]  & Hybrid-CNN        & \begin{tabular}{@{}c@{}}Energy Theft \&\\detection\end{tabular}   & AMI      & Python \\
\hline
[11]  & DL+WSVD        & \begin{tabular}{@{}c@{}}DC microgrid cyber\&\\attack detection\end{tabular}   & \begin{tabular}{@{}c@{}}DC\\Microgrid\end{tabular}      & \begin{tabular}{@{}c@{}}MATLAB/\\Python\end{tabular}\\
\hline
[12]  & GBDT (XGBoost)        & \begin{tabular}{@{}c@{}}Anomaly \&\\classification\end{tabular}   & Distribution      & Python \\
\hline
[14]  & KD+FL        & \begin{tabular}{@{}c@{}}Edge \&\\Intelligence\end{tabular}   & Alot/CPS      & Python \\
\hline
\end{tabular}
\vspace{2mm}

\begin{minipage}{\columnwidth}
\noindent
\justifying
\fontsize{7pt}{8pt}\selectfont
GBDT -- Gradient Boosted Decision Trees, AIoT -- Artificial Intelligence + Internet of Things,
AMI -- Advanced Metering Infrastructure, WSVD -- Wavelet-based Singular Value Decomposition,
SSL -- Semi-Supervised Learning.
\end{minipage}

\end{table}

\section{Virtual Microgrid AND Cyberattack Injection Architecture}
\subsection{Virtual Microgrid Overview}
\begin{figure}[ht]
    \centering
    \includegraphics[width=\linewidth]{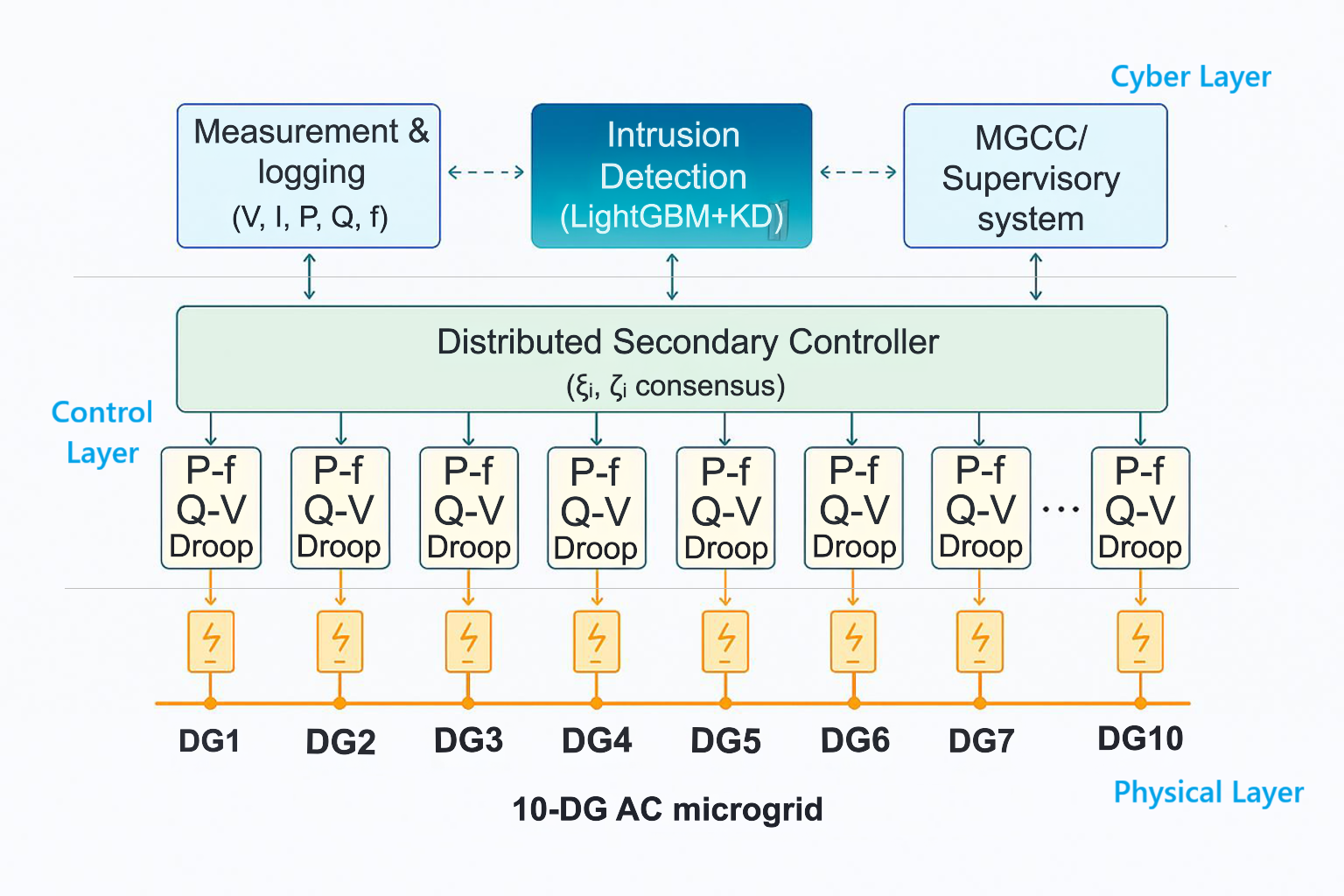}
    \caption{Architecture of the 10-DG virtual AC microgrid, showing the cyber layer, distributed secondary controller ($\xi_i$, $\zeta_i$ consensus), and local P--f / Q--V droop controllers.}
    \label{fig:microgrid_architecture}
\end{figure}

A detailed AC microgrid model was created in MATLAB/Simulink to represent a low-voltage distribution system with several inverter-based distributed generators (DGs). The structure follows well-known cyber–physical microgrid designs, such as the PSERC resiliency model by Srivastava et al. \cite{srivastava2020microgrid} and the cyber-secure distributed secondary-control architecture described by Rath et al \cite{rath2020cyber}. As shown in Fig.~\ref{fig:microgrid_architecture}, the model is expanded to include more DG units so that it can generate richer data for training machine-learning intrusion-detection models.
The electrical system contains ten three-phase voltage-source inverters. Each inverter includes an LC output filter, feeder impedance, and standard current and voltage control loops implemented in the synchronous \textit{dq} reference frame. Primary \textit{P–f} and \textit{Q–V} droop controllers manage power sharing among the DGs, while a distributed secondary controller restores the system’s voltage and frequency to their reference values. This layered control structure is consistent with widely used microgrid control approaches \cite{mahmoud2015control}, \cite{kundur2004definition}.
The cyber layer receives measurements such as bus voltages 
\textit{$V_1$, $V_2$, $V_3$}, line currents \textit{$I_1$, $I_2$, $I_3$}, and each DG's active power \textit{$P_{\mathrm{DGi}}$}, reactive power \textit{$Q_{\mathrm{DGi}}$}, and estimated frequency \textit{$f_{\mathrm{DGi}}$}, for \textit{i=1,…,10}. These measurements act both as inputs to the controllers and as features for the intrusion-detection dataset.

All simulations use a fixed step size of $2~\mu\text{s}$, producing 500,001 samples per scenario during a one-second time window. Each simulation starts in normal operating conditions until the microgrid reaches steady state; afterward, different cyberattacks are applied to the secondary-control channels. The resulting time-series data of voltages, currents, real and reactive power, and frequencies \textit{(V,I,P,Q,f)} are saved in Comma-separated values (CSV) format. Each dataset contains the following columns:

\textit{time, $V_1$, $V_2$, $V_3$, $I_1$, $I_2$, $I_3$, $P_{\mathrm{DG1}}$, $Q_{\mathrm{DG1}}$, $f_{\mathrm{DG1}}$, \dots, 
$P_{\mathrm{DG10}}$, $Q_{\mathrm{DG10}}$, $f_{\mathrm{DG10}}$, label}

The label column shows whether the system is in normal operation or experiencing a specific type of attack. These high-resolution multivariate signals form the virtual sensor dataset used to train the machine-learning models described in Section IV.

\subsection{Primary and Secondary Control Architecture}
The microgrid operates under a hierarchical control structure with two main layers, which include primary droop control at each DG and a distributed secondary controller that restores the voltage and frequency to their nominal values.
This structure is based on the PSERC resiliency framework described by Srivastava et al. \cite{srivastava2020microgrid} and the cyber-secure distributed control architecture proposed by Rath et al. \cite{rath2020cyber}.

\begin{figure}[ht]
    \centering
    \includegraphics[width=\linewidth]{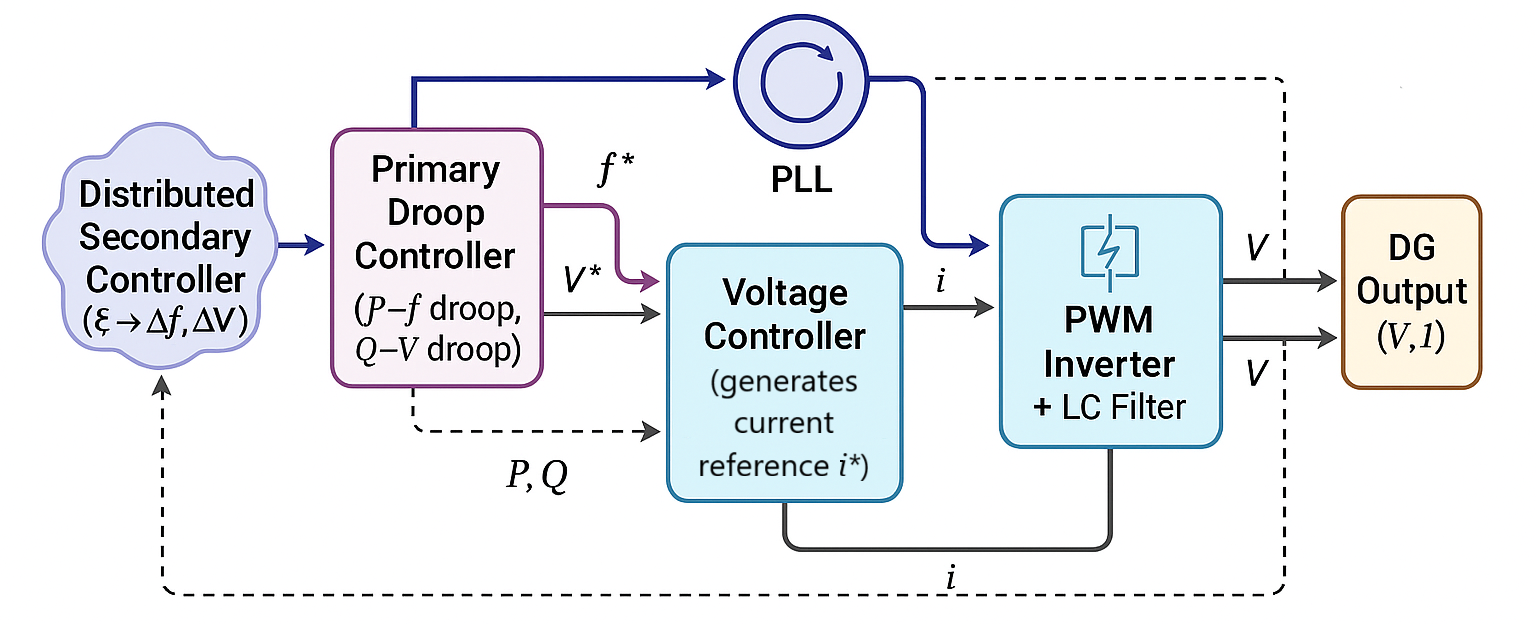}
    \caption{Cascaded control architecture of an inverter-based distributed generator (DG) unit, showing secondary control inputs, droop regulation, PLL synchronization, voltage and current control loops, and PWM inverter interface.}
    \label{fig:cascaded_control}
\end{figure}

In this work, the architecture is extended to support ten DGs and the cyberattack scenarios considered in our study.
Fig.~\ref{fig:cascaded_control} shows the step-by-step control setup for each DG, where the secondary controller adjusts the droop settings, the PLL keeps the system in sync, and the voltage and current controllers create the Pulse-width modulation (PWM) signals that control the DG’s output.
\subsubsection{Primary Droop Control}
Each inverter-based DG uses standard \textit{P–f} and \textit{Q–V} droop control to share load without fast communication links \cite{srivastava2020microgrid}.
For DG \textit{i}, the droop equations are:
\begin{equation}
\omega_i = \omega^\ast - m_i (P_i - P_i^\ast),
\label{eq:droop_freq}
\end{equation}
\begin{equation}
V_i = V^\ast - n_i (Q_i - Q_i^\ast),
\label{eq:droop_volt}
\end{equation}
where $\omega^\ast$ and $V^\ast$ are the nominal frequency and voltage,\\  
$P_i$ and $Q_i$ are the measured real and reactive power,\\  
$P_i^\ast$ and $Q_i^\ast$ are the power setpoints,\\  
and $m_i$ and $n_i$ are the droop coefficients.\\
\\
Droop control enables proportional power sharing, but it causes small steady-state errors in voltage and frequency, which motivates the need for secondary control \cite{srivastava2020microgrid}.
\subsubsection{Distributed Secondary Control}
To eliminate the steady-state errors caused by droop control, each DG participates in a distributed consensus-based secondary controller. This approach follows the method introduced in \cite{rath2020cyber}.
For DG \textit{i}, the frequency-correction dynamics are:
\begin{equation}
\dot{\xi}_i = -k_p(\omega_i - \omega^\ast) - c_f \sum_{j \in N_i} a_{ij} (\xi_i - \xi_j),
\label{eq:secondary_freq_dyn}
\end{equation}

\begin{equation}
\omega_i^{\text{cmd}} = \omega_i + \xi_i,
\label{eq:secondary_freq_cmd}
\end{equation}
and the voltage-correction dynamics are:
\begin{equation}
\dot{\zeta}_i = -k_q(V_i - V^\ast) - c_v \sum_{j \in N_i} a_{ij} (\zeta_i - \zeta_j),
\label{eq:secondary_voltage_dyn}
\end{equation}

\begin{equation}
V_i^{\text{cmd}} = V_i + \zeta_i.
\label{eq:secondary_voltage_cmd}
\end{equation}
where, $\xi_i$ and $\zeta_i$ are integral correction states,\\  
$N_i$ is the communication neighborhood of DG $i$,\\  
$a_{ij}$ are the adjacency matrix weights,\\  
and $k_p$, $k_q$, $c_f$, and $c_v$ are the secondary-control gains.\\
These equations follow the distributed controller in \cite{rath2020cyber} but are scaled to the ten-DG microgrid used in this study.
\subsubsection{Communication Layer (Attack Target)}
The secondary controller depends on low-bandwidth communication links between DGs.
As described in \cite{rath2020cyber}, the following variables are shared over the network and are vulnerable to cyberattacks:
\begin{itemize}
    \item measured frequency $\omega_i$ and voltage $V_i$,
    \item secondary correction terms $\xi_i$ and $\zeta_i$,
    \item DG-to-DG synchronization signals,
    \item feedback messages exchanged with the microgrid central controller (MGCC).
\end{itemize}
These communication channels form the main attack surface for the scenarios described later in Section 4.
\subsubsection{Inner Voltage and Current Loops}
Below the droop and secondary layers, each DG uses PI-based controllers in the dq-frame \cite{srivastava2020microgrid}.
The voltage loop is:
\begin{equation}
v_d^\ast =
V_i^{\text{cmd}} + K_{pv}(V_d - V_d^\ast) + K_{iv} \int (V_d - V_d^\ast), dt,
\label{eq:voltage_loop}
\end{equation}

and the current loop is:
\begin{equation}
i_d^{\text{cmd}} =
I_d^\ast + K_{pc}(i_d - i_d^\ast) + K_{ic} \int (i_d - i_d^\ast), dt.
\label{eq:current_loop}
\end{equation}

These inner loops are not directly attacked, but convert corrupted secondary-control signals into physical disturbances that appear in voltage, current, and power waveforms.

\subsection{Secondary Control Interface for Attack Injection}\label{sec control}
The secondary control layer is the most communication-dependent component of the microgrid architecture, making it a natural target for cyberattacks. In this work, all malicious signals are injected directly into the secondary control input channels, specifically the distributed consensus signals $\xi_i$ and $\zeta_i$ that restore frequency and voltage across all distributed generators. 
\begin{itemize}
    \item Attacks on $\xi_i$ primarily distort active power and frequency regulation,
    \item Attacks on $\zeta_i$ impact reactive power and voltage regulation,
    \item Attacks on both signals emulate coordinated disturbances affecting overall stability.
\end{itemize}

\begin{figure}[ht]
    \centering
    \includegraphics[width=\linewidth]{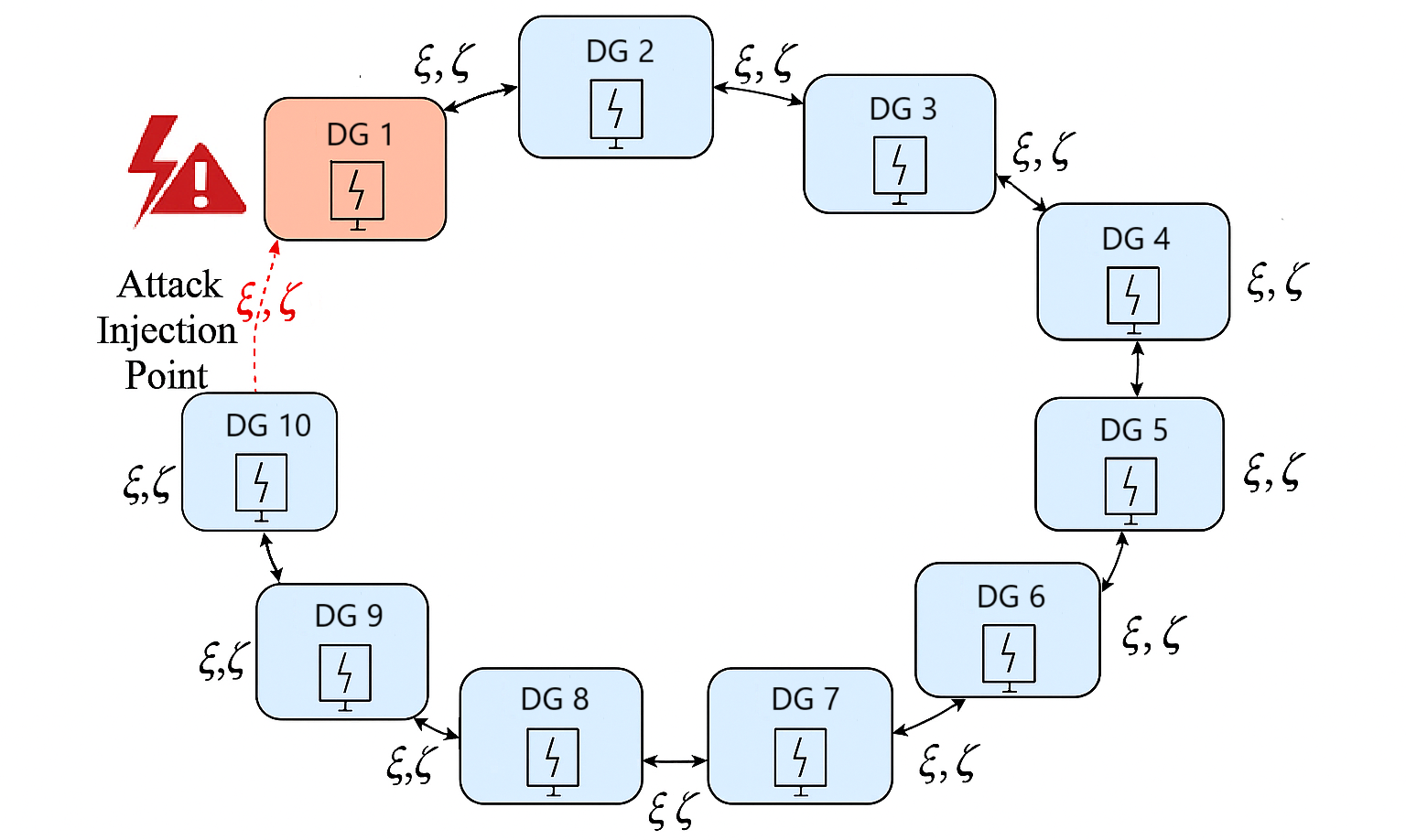}
    \caption{Communication Pathways of the Secondary Control Layer With the Injection Point of the Corrupted Consensus Signals.}
    \label{fig:Consensus_based}
\end{figure}

Fig.~\ref{fig:Consensus_based} shows the ring-based secondary-control communication network used by the ten inverter-based DG units. Each DG exchanges the consensus signals $\xi$ and $\zeta$ with its neighbors, and the attack is injected by corrupting the incoming signal received by DG1.
This design choice follows the cyber-physical vulnerability analysis presented by Srivastava et al. \cite{srivastava2020microgrid} and the threat model from the cyber-secure microgrid architecture paper used in this work. 
Because the DGs share information only with their neighbors over a sparse communication network, an attacker who breaks into just one communication link can affect several DGs at once. Earlier research has shown that even small false signals added to the secondary controller can spread through the network, change how power is shared, and weaken the microgrid’s dynamic response \cite{huang2024impact,beg2023review}.

\begin{figure}[ht]
    \centering
    \includegraphics[width=\linewidth]{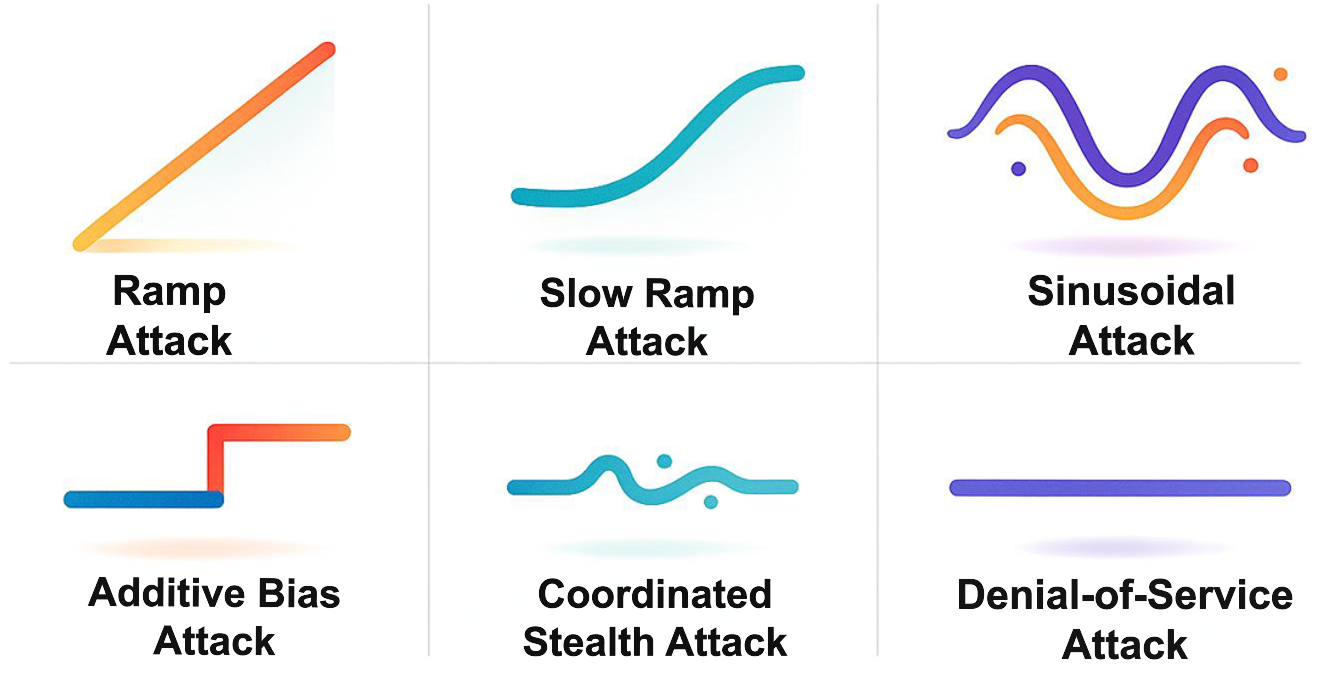}
    \caption{Overview of the six secondary-control cyberattack patterns applied to the consensus signal  $\xi_i (t)$ and $\zeta_i(t)$.}
    \label{fig:Overview_six_control}
\end{figure}

To study these weaknesses under realistic conditions, this work applies six types of cyberattacks. Each attack is injected into the secondary-control input of selected DGs. Fig.~\ref{fig:Overview_six_control} provides a compact illustration of the six secondary-control attack profiles considered, highlighting the characteristic shape of each injected disturbance.

\subsubsection{Additive Bias Attack}
A constant malicious offset is added to both secondary signal, which is given by:
\begin{align}
\xi_i^{\text{mal}}(t) &= \xi_i(t) + b, \label{eq:eq13}\\
\zeta_i^{\text{mal}}(t) &= \zeta_i(t) + b. \label{eq:eq14}
\end{align}
This attack shifts the internal reference points of the controller, producing sustained misregulation in both active- and reactive-power sharing. Fixed-bias spoofing is a classical false-data injection approach and is rigorously analyzed in the foundational work of the literature \cite{liu2011false}.

\subsubsection{Ramp Attack}
A gradually increasing bias is added to the distributed secondary-control frequency-correction signal $\xi_i$. The malicious modification is expressed as:
\begin{equation}
\xi_i^{\text{mal}}(t) = \xi_i(t) + r t
\label{eq:eq9}
\end{equation}
where \text{r} is the ramp slope.
Because the frequency loop is very sensitive to slow long-term drift, even small errors can gradually build up and create an active-power imbalance. By imitating a slowly increasing load, the attacker causes the controller to shift power incorrectly. Similar slow-moving biases have been shown to destabilize secondary control loops in microgrids when not corrected \cite{beg2023review, liu2022efficient,liu2011false}.

\subsubsection{Slow Ramp Attack}
A very small slope is injected into the secondary voltage-restoration signal. The attack modification is given by:
\begin{equation}
\zeta_i^{\text{mal}}(t) = \zeta_i(t) + r_s t,\quad r_s \ll r. \label{eq:eq10}
\end{equation}
This attack targets $\zeta_i$ because the voltage loop can conceal very slow deviations, allowing the attacker to create a stealthy reactive-power imbalance without triggering threshold-based detectors. Such slowly drifting anomalies have been reported in voltage stability assessments and cyber-physical intrusion analyses \cite{kundur2004definition,beg2023review}.

\subsubsection{Sinusoidal Attack}
A coordinated oscillatory signal is injected into both secondary control inputs. The malicious change is given by;
\begin{align}
\xi_i^{\text{mal}}(t) &= \xi_i(t) + A \sin(\omega_a t), \label{eq:eq11}\\
\zeta_i^{\text{mal}}(t) &= \zeta_i(t) + A \sin(\omega_a t). \label{eq:eq12}
\end{align}
Oscillatory disturbances affect both P and Q regulation, exciting inverter dynamics and distorting transient responses. Injecting them simultaneously into $\xi_i$ and $\zeta_i$ emulates a coherent, dynamic intrusion, similar to oscillatory false-data injections shown to degrade microgrid performance \cite{dehghani2021cyber}.

\subsubsection{Coordinated Stealth Attack}
In this mode, small, correlated deviations are injected across multiple DGs.

\begin{align}
\xi_i^{\text{mal}}(t) &= \xi_i(t) + f_s(t), \label{eq:eq15}\\ \zeta_i^{\text{mal}}(t) &= \zeta_i(t) + \alpha f_s(t). \label{eq:eq16} \end{align}

where $ f_s(t)$ is a low-amplitude waveform designed to remain within normal envelopes.\\
This mode targets both control loops because coordinated multi-variable manipulation is one of the most difficult patterns to detect. By maintaining correlations between the disturbances, the attack closely mimics normal operation while degrading power-sharing performance. Similar multi-signal stealth strategies appear in recent intrusion-detection literature\cite{qi2021detecting,gunduz2024smart}.
\subsubsection{Denial-of-Service (DoS) Attack}
A DoS attack freezes the frequency-restoration signal by preventing new consensus data from reaching DGs. This malicious modification is given by:
\begin{equation}
\xi_i^{\text{mal}}(t) = \xi_i(t_k), \qquad t > t_k .
\label{eq:eq17}
\end{equation}
The frequency channel is targeted because loss of synchronization typically manifests first through degraded P–f restoration. With stale control data, DGs drift apart in frequency and voltage, leading to instability. DoS problems in microgrid control are well documented in the literature, with similar observations reported in broader cybersecurity surveys \cite{beg2023review,mahmoud2015control}.

All attacks were injected exclusively into the secondary-control correction layer, ensuring that disturbances propagate through the cyber-control channel rather than through sensors or inverter hardware. This modeling approach aligns with real microgrid vulnerabilities documented in recent literature \cite{srivastava2020microgrid,beg2023review,mahmoud2015control}.
All simulations were run in MATLAB/Simulink using MGLib within the custom virtual microgrid built for this study.
\subsection{MATLAB/Simulink Implementation and Data Logging Framework}
The virtual microgrid and cyberattack-injection environment were implemented entirely in MATLAB/Simulink using custom Simscape-based inverter models, hierarchical controllers, and structured cyberattack subsystems. Two simulation environments, Normal.slx and MGLib.slx, were used. Normal.slx implements the full 10-DG microgrid and executes all cyberattack scenarios, while MGLib.slx provides a library of reusable distributed-generation and control modules used to construct and parameterize the virtual microgrid. The modeling philosophy follows the cyber-physical microgrid resilience principles and the distributed secondary control architecture outlined in the literature \cite{srivastava2020microgrid,liu2011false}.
\subsubsection{Model Structure}
As illustrated in Fig.~\ref{fig:microgrid_architecture}, the Simulink model follows the physical, control, and cyber–physical behavior of an inverter-dominated microgrid and is organized into four connected layers.
\subsubsection*{a) Physical Layer}
The physical layer contains all inverter-based DG units, along with the AC distribution lines and system loads. Each DG uses a voltage-source inverter model with droop-based primary control, following the standard methodology presented in \cite{mahmoud2015control}. The switching behavior is fully retained, allowing realistic harmonic and dynamic responses.
\subsubsection*{b) Secondary Control Layer}
A distributed consensus-based voltage and frequency restoration controller regulates the network. Each DG exchanges correction signals $\xi_i$ and $\zeta_i$  with its neighbors. The controller setup is based on the cyber-secure distributed control framework introduced in \cite{rath2020cyber}, which describes how communication links, coordination graphs, and vulnerability paths operate in autonomous AC microgrids.
\subsubsection*{c) Cyber Layer (Communication Graph)}
Peer-to-peer communication links between DGs are implemented using Simulink signal routing and discrete update blocks. Communication irregularities such as delay, sample jitter, and packet manipulation are introduced at this layer. These imperfections reflect realistic cyber-physical interactions, consistent with observations from recent microgrid cybersecurity reviews \cite{akinwale2021mitigation}.
\subsubsection*{d) Attack-Injection Layer}
This layer contains structured cyberattack blocks that inject ramps, biases, oscillations, or packet-freeze events directly into the control corrections $\xi_i$ and $\zeta_i$. The attack signals follow the formulations defined in Section \ref{sec control} and are designed to propagate through the entire secondary-control loop in a physically meaningful manner.
\subsubsection{Attack Injection Blocks}
Each of the six cyberattacks defined in Section III-B was implemented as an independent Simulink subsystem so that scenarios can be executed in a controlled and repeatable way. The subsystems realize the ramp, slow-ramp, sinusoidal, additive-bias, coordinated stealth, and DoS profiles using standard integrator, sine-wave, offset, and sample-and-hold blocks following the formulations in (\ref{eq:eq9})--(\ref{eq:eq17}).\\
This set of attacks follows the cyber-physical disturbance pathways described in \cite{srivastava2020microgrid} and the microgrid cyberattack case studies summarized by \cite{beg2023review}.
\subsubsection{High-Frequency Simulation Parameters}
All simulations were executed using high-fidelity switching resolution:
\begin{itemize}
    \item Sample time: $2~\mu\text{s}$
    \item Solver: Fixed-step, discrete
    \item Switching model: High-resolution PWM behavior
    \item Simulation duration: 1.0 s
    \item Step size: $2 \times 10^{-6} \text{ s}$
\end{itemize}
Such small time steps follow best practices for simulating inverter-dominated microgrids. They allow the model to capture switching ripple and fast system dynamics accurately, which agrees with commonly accepted transient-modeling guidelines \cite{kuyumcu2025high,guruwacharya2024data}.
\subsubsection{Data Logging and Export Pipeline}
A dedicated logging subsystem recorded all cyber-physical signals from every DG:
\begin{itemize}
    \item Voltage $V_i$, Current $I_i$, Real/reactive power $P_{\mathrm{DG1}}$, $Q_{\mathrm{DG1}}$, Frequency estimate $f_{\mathrm{DG1}}$
    \item Attack state (binary label: 0 = normal, 1 = attack)
    \item Attack type index (Normal, Ramp, Slow Ramp, Additive, Sinusoid, Stealth, DoS)
    \item Time index
\end{itemize}

The signals were saved using To Workspace and To File blocks in array form, and then exported to CSV files using MATLAB’s built-in data-export functions.

Each simulation produced about 500,001 samples across 38 variables, resulting in high-resolution datasets that are suitable for machine-learning analysis.

\subsubsection{Dataset Consolidation}
All scenario outputs (Normal and six attack types) were exported as CSV files. These files were later merged and prepared into a unified supervised-learning dataset, as detailed in Section IV.
\subsection{Noise Representation and Robustness Modeling}
\begin{figure}[ht]
    \centering   \includegraphics[width=\linewidth]{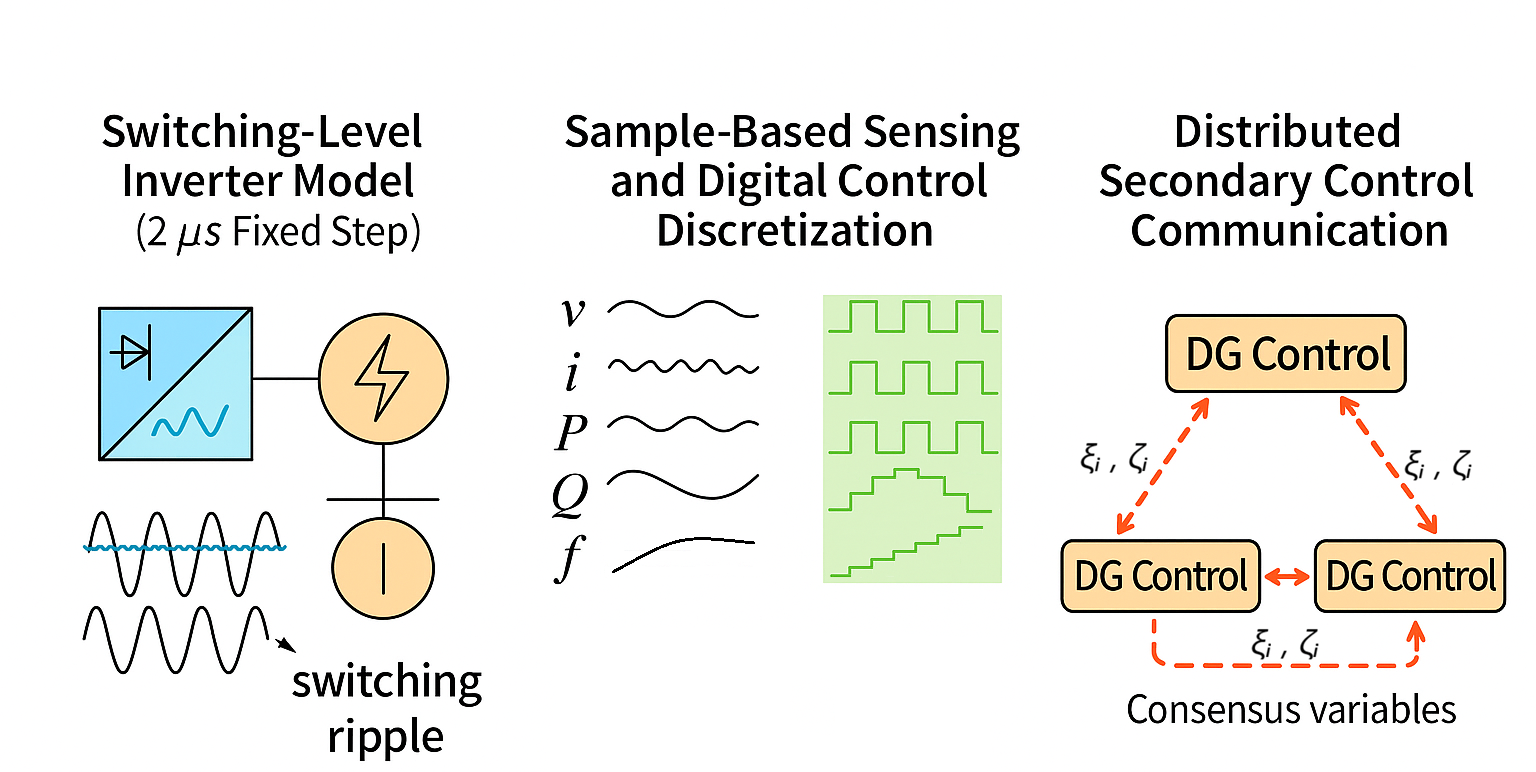}
    \caption{Illustration of noise and discretization effects in the virtual microgrid, showing PWM switching ripple in voltage and current signals, quantization steps from digital measurement, and small timing jitter in the consensus updates of the secondary controller.}
    \label{fig:Noise}
\end{figure}

Although cyberattacks disturb the secondary-control layer, inverter-based microgrids also contain natural physical and numerical noise even during normal operation. This noise comes mainly from high-frequency inverter switching, digital sampling, and discretized control updates. These effects introduce small ripples and variability into measured voltage, current, power, and frequency signals. To capture these realistic disturbances, the virtual microgrid was simulated using a switching-level inverter model with a fixed $2~\mu\text{s}$ step size. As shown in Fig.~\ref{fig:Noise}, this produces switching ripple at the electrical level, staircase-like sampled measurements at the control level, and discrete-time communication of consensus variables in the secondary controller. Together, these effects create structured noise and variability similar to what is observed in practical inverter-based microgrids~\cite{srivastava2020microgrid}. This ensures that the machine-learning models are trained on data that reflect real operating conditions rather than idealized, noise-free signals.

\subsubsection{Measurement Noise from High-Fidelity Switching Dynamics}
Each distributed generator uses a PWM-based voltage source inverter modeled at microsecond resolution. As noted in prior transient modeling studies of power-electronic systems \cite{chandorkar1993control}, switching ripple introduces small but persistent harmonic components into voltage and current waveforms \cite{guruwacharya2024data}. These distortions act as a realistic source of sensor noise and ensure that the resulting dataset reflects the non-ideal characteristics of real converter-dominated microgrids. Unlike artificially smoothed or phasor-domain models, this switching representation prevents the classifier from overfitting to unrealistically clean signals.
\subsubsection{Numerical Discretization Noise in Sensing and Control}
The microgrid uses sample-based measurement of real power, reactive power, and frequency at each DG. Because these quantities are computed from discrete-time signals, they exhibit quantization effects and numerical jitter \cite{widrow2008quantization}. Such effects mimic the finite precision and sampling behavior of real-world measurement systems and digital controllers. This type of noise is particularly important when classifying subtle or slowly varying attacks such as Slow Ramp and Coordinated Stealth disturbances, which must be detected against a background of small natural fluctuations \cite{srivastava2020microgrid}.
\subsubsection{Communication Jitter in Distributed Secondary Control}
The distributed secondary control layer operates through discrete communication exchanges among DG controllers. Although no explicit random delay blocks were added, the discrete-time update behavior of Simulink inherently produces small timing mismatches between controller nodes. As described in \cite{srivastava2020microgrid} and prior reviews of microgrid cyber-physical vulnerabilities \cite{beg2023review}, such jitter is representative of real-world low-bandwidth microgrid communication networks. This adds realistic uncertainty to consensus variables $\xi_i$ and $\zeta_i$, improving the resilience of the classifier to benign fluctuations in the absence of attacks.
\subsubsection{Implications for Machine Learning Robustness}
The combination of switching ripple, discretization noise, and communication jitter forms a realistic disturbance environment. These factors help ensure that the LightGBM models trained in later sections do not overfit to clean or idealized trajectories. Instead, they learn to classify attacks under noisy, non-linear, and physically accurate operating conditions, consistent with recommendations in prior microgrid intrusion detection studies \cite{beg2023review,dehghani2021cyber,drayer2018detection}. This contributes directly to the generalization ability shown in Section V, where the classifier achieves high accuracy across all attack modes.

\section{Dataset Generation and Machine Learning Methodology}
This section describes how high-resolution microgrid simulation data were converted into structured datasets for supervised learning, and how the machine-learning models were designed, trained, and optimized. All data originate from the MATLAB/Simulink virtual microgrid developed in Section III, where cyberattacks were injected directly into the distributed secondary-control layer.
\subsection{High-Resolution Data Generation in Simulink}
Each scenario, whether normal or under one of the six attack modes, was simulated for 1 s using a $2~\mu\text{s}$ fixed-step solver, yielding 500,001 time samples. Small time steps are recommended for inverter-dominated microgrids because they preserve fast switching behavior, harmonic content, and non-linear transient dynamics that influence cyber-physical stability, as emphasized in modern power-electronics transient modeling studies \cite{rocabert2012control,pogaku2007modeling}.
For each time step, the following measurements were logged:
\begin{itemize}
    \item Three-phase bus voltages: $V_1(t)$, $V_2(t)$, $V_3(t)$
    \item Three-phase line currents: $I_1(t)$, $I_2(t)$, $I_3(t)$
    \item Active and reactive power of $DG_{i}$, $i \in \{1,\dots,10\}$: 
$P_{\mathrm{DG}i}(t),\; Q_{\mathrm{DG}i}(t)$ 
    \item Local frequency measurements: $f_{\mathrm{DGi}}(t)$
    \end{itemize}
    These variables form a virtual sensor layer analogous to what would be measured by Phasor measurement units (PMUs), smart inverters, or microgrid controllers in practice.

Each simulation also included:
\begin{itemize}
    \item Attack state (binary label):
    $y_{\text{bin}} \in \{0~\text{(normal)},\, 1~\text{(attack)}\}$

    \item Attack type index (Normal, Ramp, Slow Ramp, Additive, Sinusoid, Stealth, DoS)
\end{itemize}
\subsection{Dataset Consolidation and Attack Annotation}
After simulation, all CSV files were merged into a single dataset.
Scenario filenames were parsed to automatically assign:
\begin{itemize}

\item Binary Label:

\[
y_{\text{bin}} =
\begin{cases}
0, & \text{normal} \\
1, & \text{attack}
\end{cases}
\]

\item {Multiclass Attack Type:}

$y_{\text{multi}} \in \{0:\text{Normal},\; 1:\text{Additive},\; 
2:\text{Ramp},\; 3:\text{Slow Ramp},\; 4:\text{Sinusoid},\; 
5:\text{Stealth},\; 6:\text{DoS}\}$

\item The combined dataset has a size of $500{,}001 \times 7 \approx 3.5$ million samples.\\
This included 38 continuous features and two target labels.
\end{itemize}

\subsection{Memory-Aware Downsampling Strategy}

Training on the full 3.5-million–sample dataset exceeded available GPU/CPU memory.
To retain fidelity while reducing size, a balanced downsampling scheme was used \cite{chen2015net2net}. All attack samples were kept, and only 10–-15\% of normal samples were retained. This achieved a 50-–60\% reduction while preserving transients at attack onset, normal–attack boundaries, and temporal dynamics essential for classification. The final working dataset contained approximately 1,030,197 samples.
\subsection{Noise-Resilient Feature Normalization}
Because the microgrid includes switching ripple, discretization noise, and communication jitter (Section III), normalization must preserve subtle variations.
To avoid memory overflow, incremental (chunk-based) normalization was adopted using the standard z-score formula \cite{lin2003symbolic}:
\begin{equation}
x_{\text{norm}} = \frac{x - \mu}{\sigma}
\label{eq:x_norm}
\end{equation}
where \( x \) denotes the raw feature value, \( \mu \) is the feature mean, and \( \sigma \) is the corresponding standard deviation.

The key points to note are that the Data were converted to float32 to reduce the memory footprint. Large numerical fields were rescaled (e.g., division by 1000) to avoid overflow, and normalization was computed in chunks to avoid Random Access Memory (RAM) exhaustion. The resulting dataset had $\mu \approx 0 \;\text{and}\; \sigma \approx 1$ for all features.

\subsection{Train–Validation–Test Split}
A stratified split was used to keep all attack types represented in the same proportions across the training, validation, and test sets \cite{buitinck2013api}. The training set contains 721,137 samples, while the validation and test sets each contain 154,530 samples. Stratification covered both Binary label and Multiclass attack type to avoid bias and maintain class balance.
\subsection{Binary Attack Detection Using LightGBM}
Binary classification separates normal operation from attack conditions. LightGBM is used because it is fast on structured data, requires very little memory, performs extremely well without Graphics processing unit (GPU) acceleration, and has a strong track record in cyber-physical system applications \cite{ke2017lightgbm}.
The model minimized the standard binary log-loss:
\begin{equation}
L_{\text{bin}} = - \left[ y \ln(\hat{p}) + (1 - y)\ln(1 - \hat{p}) \right]
\end{equation}
where $y \in \{0,1\}$ is the ground--truth binary label (0 = normal, 1 = attack) and 
$\hat{p}$ is the predicted probability that the sample belongs to the attack class.

\subsection{Multiclass Attack-Type Classification}

A second LightGBM model was trained to classify all 7 attack types at the same time. The goal was for the model to correctly tell the difference between each type of cyberattack. To achieve this, the model minimized the standard multiclass cross-entropy (log-loss) function:
\begin{equation}
L_{\text{mc}} = - \sum_{k=1}^{7} y_k \ln\left(\hat{p}_k\right)
\end{equation}

where $y_k$ is the one--hot encoded ground--truth label and $\hat{p}_{\,k}$ is the predicted probability for class $k$.
The confusion matrix (Section V) confirms nearly perfect class separability.
\subsection{Knowledge Distillation for Lightweight Deployment}
To reduce inference time and model size, knowledge distillation (KD) was used \cite{hinton2015distilling}.

A teacher–student architecture was implemented:
\begin{itemize}
\item Teacher: full LightGBM multiclass model

\item Student: smaller LightGBM model trained using Hard labels, Soft probabilities from the teacher, and Temperature-scaled softmax.
\end{itemize}
\[
L = \alpha \, CE(y_{\text{hard}}, s) 
    + \beta \, KL\!\left( \mathrm{softmax}(T_t),\, \mathrm{softmax}(T_s) \right)
\]
where:
\begin{itemize}
\item $t$ and $s$ are the teacher and student logits,
\item $T$ is the distillation temperature,
\item $CE$ is the standard cross-entropy loss using hard labels,
\item $KL$ is the Kullback--Leibler divergence between teacher and student softened probabilities,
\item $\alpha$ and $\beta$ control the weight of the hard-label and soft-label losses.
\end{itemize}.

The LightGBM-based architecture meets real-time requirements and achieves efficiency gains consistent with findings in \cite{ke2017lightgbm, gou2021knowledge}.

\section{Results}

This section evaluates the proposed cyberattack-detection framework. It first examines how secondary-control attacks appear in physical measurements, and then reports the performance of the machine-learning models. The results cover binary attack detection, multiclass attack classification, feature-importance analysis, knowledge-distillation outcomes, and real-time prediction behavior. All experiments were performed on the normalized, downsampled dataset produced from the high-resolution simulations described in Section IV.

\subsection{Physical Signatures of Secondary-Control Attacks}

\begin{figure}[ht]
    \centering
    \includegraphics[width=\linewidth]{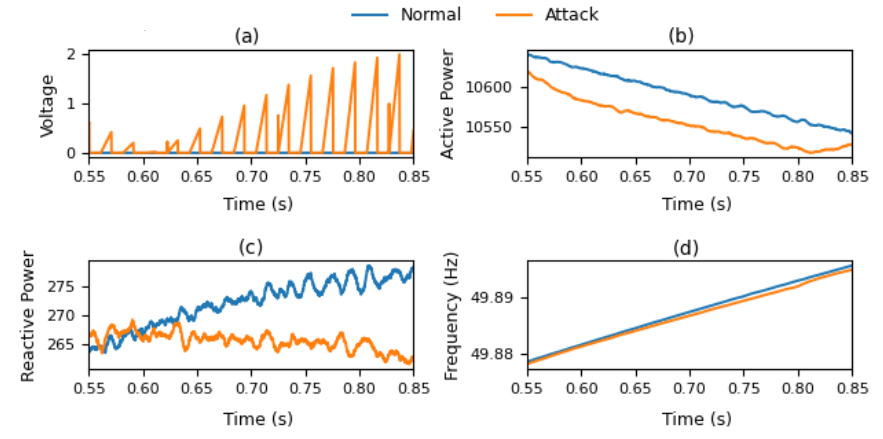}
    \caption{Representative DG1 time-domain response under an additive-bias attack injected at $t_a=0.70$ s. The attack perturbs the secondary-control consensus signals, producing measurable deviations in (a) bus voltage $V_1(t)$, (b) active power $P_{\mathrm{DG1}}(t)$, (c) reactive power $Q_{\mathrm{DG1}}(t)$, and (d) frequency estimate $f_{\mathrm{DG1}}(t)$ compared with normal operation.}
    \label{fig:Additive}
\end{figure}

To show how secondary-control attacks appear in physical measurements, Fig.~\ref{fig:Additive} shows the time-domain response of DG1 under an additive-bias attack. The figure compares normal and attacked values of bus voltage $V_1(t)$, active power $P_{\mathrm{DG1}}(t)$, reactive power $Q_{\mathrm{DG1}}(t)$, and frequency $f_{\mathrm{DG1}}(t)$.

Even over a short time window, the additive attack causes visible changes in all four variables. The attacked curves slowly move away from the normal ones, showing that a bias added at the secondary-control layer passes through the droop controller and inner control loops and appears in the electrical outputs of DG1. Changes in $P_{\mathrm{DG1}}(t)$ and $Q_{\mathrm{DG1}}(t)$ show that power sharing is altered, while deviations in $V_1(t)$ and $f_{\mathrm{DG1}}(t)$ show weaker voltage and frequency regulation. Although the change in any single variable may look small, the combined behavior of $(V, P, Q, f)$ forms a clear pattern that can be used for attack detection.

\begin{figure}[ht]
    \centering
    \includegraphics[width=\linewidth]{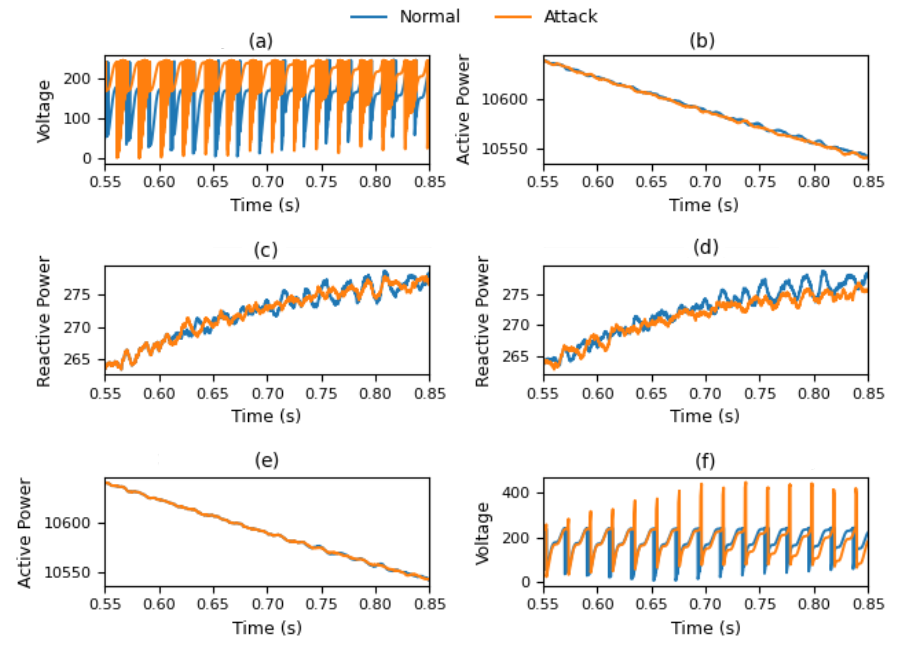}
    \caption{Representative DG1 time-domain responses under different secondary-control cyberattacks compared with normal operation. 
(a) Ramp attack: bus voltage $V_1(t)$. 
(b) Ramp attack: active power $P_{\mathrm{DG1}}(t)$. 
(c) Slow-ramp attack: reactive power $Q_{\mathrm{DG1}}(t)$. 
(d) Sinusoidal attack: reactive power $Q_{\mathrm{DG1}}(t)$. 
(e) Coordinated stealth attack: active power $P_{\mathrm{DG1}}(t)$. 
(f) Denial-of-service (DoS) attack: bus voltage $V_1(t)$. 
Each panel compares the normal (blue) and attacked (orange) trajectories.}
    \label{fig:attack_signatures}
\end{figure}

Fig.~\ref{fig:attack_signatures} compares the effects of several secondary-control attacks on DG1.

Ramp attacks cause growing distortion in $V_1(t)$ and a steady drift in $P_{\mathrm{DG1}}(t)$, showing gradually changing power sharing. Slow-ramp attacks create a small but persistent bias in $Q_{\mathrm{DG1}}(t)$. Sinusoidal attacks add oscillations and a clear offset to $Q_{\mathrm{DG1}}(t)$. Coordinated stealth attacks cause only small changes in $P_{\mathrm{DG1}}(t)$, and the attacked curve stays close to the normal one. This shows how low-amplitude coordinated disturbances can look like normal operation while still degrading performance. DoS attacks produce larger oscillations and peaks in $V_1(t)$ because frozen or delayed consensus updates weaken voltage regulation.

Overall, different secondary-control attacks leave different physical patterns in $(V, P, Q)$. Ramp and DoS attacks create clearly visible changes, while slow-ramp and stealth attacks cause small but consistent shifts. These combined changes across multiple variables provide useful features for the LightGBM classifier to detect attacks and identify their type.

\subsection{Binary Attack Detection Performance}

Using the label definitions $y_{\text{bin}}$ from Section IV-B, the binary LightGBM classifier was trained on the merged dataset to discriminate between normal and malicious operating modes. After training with early stopping, the binary model achieved the results shown in table\ref{tab:model_performance}. This shows that the model can dependably identify whether the microgrid is operating normally or under attack, even when exposed to a wide range of disturbance scenarios.

\begin{figure}[ht]
    \centering
    \includegraphics[width=\linewidth]{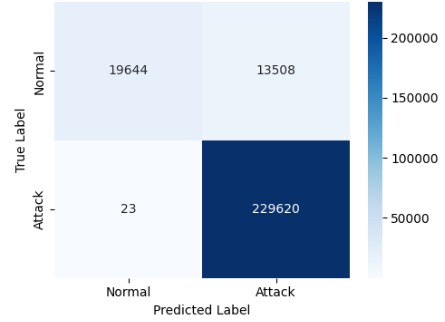}
    \caption{Binary Confusion Matrix for Cyberattack Detection in the Microgrid}
    \label{fig:Bin_Confusion}
\end{figure}

\subsubsection*{Confusion Matrix Analysis}

From Fig.~\ref{fig:Bin_Confusion}, the binary confusion matrix demonstrates:
\begin{itemize}
    \item Very low false-negative rates, meaning almost all attacks were detected.
    \item Most errors correspond to false positives, where normal operating conditions are occasionally classified as attacks.
    \item Attack samples are identified correctly in more than 99\% of cases, showing that the model stays reliable even when noise and inverter switching ripple are present.
\end{itemize}

Overall, the binary classifier works well as a strong first line of defense. It can quickly detect when a cyberattack is happening with very high sensitivity.

\subsection{Multiclass Attack-Type Classification} 

To identify specific attack types, a multiclass LightGBM model was trained using the label definitions $y_{\text{multi}}$ from Section IV-B.

\subsubsection{Test-Set Performance}

These results of the multiclass classifier, as summarized in table\ref{tab:model_performance}, indicate almost perfect classification across all attack types.

\begin{table}[t]
\centering
\caption{Performance Comparison of Binary, Multiclass, and Distilled Models}
\label{tab:model_performance}
\begin{tabular}{lccccc}
\hline
\textbf{Model} &
\textbf{Accuracy (\%)} &
\textbf{\begin{tabular}{@{}c@{}}Macro\\F1 (\%)\end{tabular}} &
\textbf{\begin{tabular}{@{}c@{}}Weighted\\F1 (\%)\end{tabular}} &
\textbf{\begin{tabular}{@{}c@{}}Inference\\(ms/1000)\end{tabular}} \\
\hline
\begin{tabular}{@{}c@{}}Binary\\LightGBM\end{tabular}     & 94.8  & 94.3  & 94.3  & 53.96 \\
\begin{tabular}{@{}c@{}}Multiclass\\LightGBM\end{tabular} & 99.72 & 99.62 & 99.72 & 67.00 \\
Student (KD)                                              & 99.70 & 99.70 & 99.70 & $\sim$18.0 \\
\hline
\end{tabular}
\end{table}

\subsubsection{Confusion Matrix Interpretation}
From Fig.~\ref{fig:Confusion}, the multiclass confusion matrix shows:
\begin{figure}[ht]
    \centering
    \includegraphics[width=\linewidth]{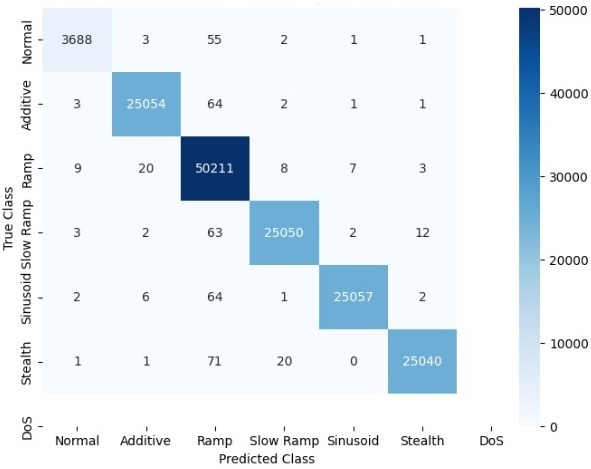}
    \caption{Multiclass Confusion Matrix for Cyberattack Detection in the Microgrid}
    \label{fig:Confusion}
\end{figure}

\begin{itemize}
    \item Strong diagonal dominance, meaning the model is very good at telling each attack type apart.
    \item Very few mistakes, even between Ramp and Slow-Ramp attacks, which normally look similar.
    \item Perfect or near-perfect recognition of Additive, Sinusoid, Stealth, and DoS attacks.
\end{itemize}

The strong separation between attack types shows that the dataset provides enough unique signals for each mode. This also shows that the model can clearly recognize these differences in the data and use them to classify each attack type accurately.

\subsection{LightGBM Hyperparameters}
The binary and multiclass classifiers were implemented using LightGBM. All model configurations, including the boosting setup, learning parameters, and iteration counts, are summarized in Table~\ref{tab:hyperparams}. The binary model employed a standard binary objective, while the multiclass model used a seven-class formulation with early stopping determining the final number of boosting rounds. A compact student model was also trained using knowledge distillation. As shown in Table~\ref{tab:hyperparams}, the student configuration uses reduced tree complexity and fewer boosting rounds to enable lower memory usage and faster inference while maintaining strong predictive performance.

\begin{table}[t]
\caption{LightGBM Hyperparameters for Binary, Multiclass, and Student Models}
\label{tab:hyperparams}
\centering
\setlength{\tabcolsep}{4pt}
\renewcommand{\arraystretch}{1.1}
\begin{tabular}{lccc}
\hline
\textbf{Parameter} & \textbf{Binary} & \textbf{Multiclass} & \textbf{Student} \\
\hline
Objective              & binary      & multiclass      & multiclass \\
Number of classes      & --          & 7               & 7 \\
Boosting type          & gbdt        & gbdt            & gbdt\footnotemark[1] \\
Num.\ leaves           & 63          & 63              & 15 \\
Learning rate          & 0.05        & 0.05            & 0.10 \\
Feature fraction       & 0.9         & 0.9             & 0.8 \\
Bagging fraction       & 0.8         & 0.8             & 0.8 \\
Bagging frequency      & 5           & 5               & --\footnotemark[1] \\
Num.\ iterations (set) & 200         & 200             & 50 \\
Best iteration         & 200         & 104             & 50 \\
Threads (\texttt{n\_jobs}) & $-1$    & $-1$            & --\footnotemark[1] \\
\hline
\end{tabular}
\footnotetext[1]{Parameters not explicitly set for the student model use LightGBM default values.}
\end{table}

\subsection{Statistical Robustness and Ablations}

Robustness was assessed using per-class F1-scores and targeted feature-removal tests. The Normal mode remains the most difficult to classify (F1 $\approx$ 0.983), while all attack classes exceed 0.996, indicating stable performance even under subtle secondary-control disturbances. As summarized in Table~\ref{tab:ablation}, removing active-power, reactive-power, or frequency signals produces negligible changes in macro-F1 (0–0.01\% drop), showing that these quantities are highly redundant across distributed generators. In contrast, removing only the three bus-voltage measurements (V1–V3) decreases macro-F1 by approximately 1 percentage point, identifying voltage as the most informative feature group for distinguishing normal and ramp-type behaviors. Line-current removal shows no measurable effect. Overall, the classifier remains highly robust, with voltage magnitudes being the only feature group that provides uniquely discriminative information for this attack set.

\begin{table}[t]
\centering
\caption{Ablation Study: Effect of Removing Each Feature Group}
\label{tab:ablation}
\begin{tabular}{lccp{2.2cm}}
\toprule
\textbf{Feature Removed} & \textbf{F1} & \textbf{Drop} & \textbf{Interpretation} \\
\midrule
None (Baseline) & 0.9948 & -- & Full feature set. \\

$P_{\mathrm{DG}}$ (Active Power) & 0.9948 & 0.00\% & No measurable impact. \\

$Q_{\mathrm{DG}}$ (Reactive Power) & 0.9948 & 0.00\% & Redundant information. \\

$f_{\mathrm{DG}}$ (Frequency) & 0.9947 & 0.01\% & Minor contribution. \\

$V_1$--$V_3$ (Bus Voltage) & 0.9847 & 1.01\% & Most informative group. \\

$I_1$--$I_3$ (Line Currents) & 0.9948 & 0.00\% & Redundant with others. \\
\bottomrule
\end{tabular}
\end{table}

\subsection{Feature-Importance Analysis}
Gain-based feature importance was computed for the multiclass classifier to understand which physical variables contribute most to attack detection.

Key observations include:
\begin{itemize}
\item Active power measurements $P_{\mathrm{DG}i}$ were dominant predictors, which aligns with the fact that many attacks influence power-sharing dynamics.

\item DG local frequencies $F_{\mathrm{DG}i}$ were critical for detecting oscillatory disturbances (Sinusoid) and synchronization-affecting attacks (Stealth, DoS).

\item Line currents $I_1$, $I_2$, $I_3$ showed high importance for detecting abrupt disturbances such as DoS and Ramp.

\item Voltage measurements played a consistent but moderate role across all classes.
\end{itemize}

\subsubsection*{Interpretation}

The feature-importance distribution indicates that the classifier relies on microgrid-wide behavioral signatures, not isolated measurements. Because it depends on many variables, it is more resilient and less likely to be misled by an attacker manipulating a single sensor.

\subsection{Knowledge Distillation Evaluation} 

KD process was used to compress the multiclass LightGBM model into a lighter student model suitable for real-time embedded controllers \cite{hinton2015distilling}.
\subsubsection*{Performance Comparison}

\begin{figure}[ht]
    \centering
    \includegraphics[width=\linewidth]{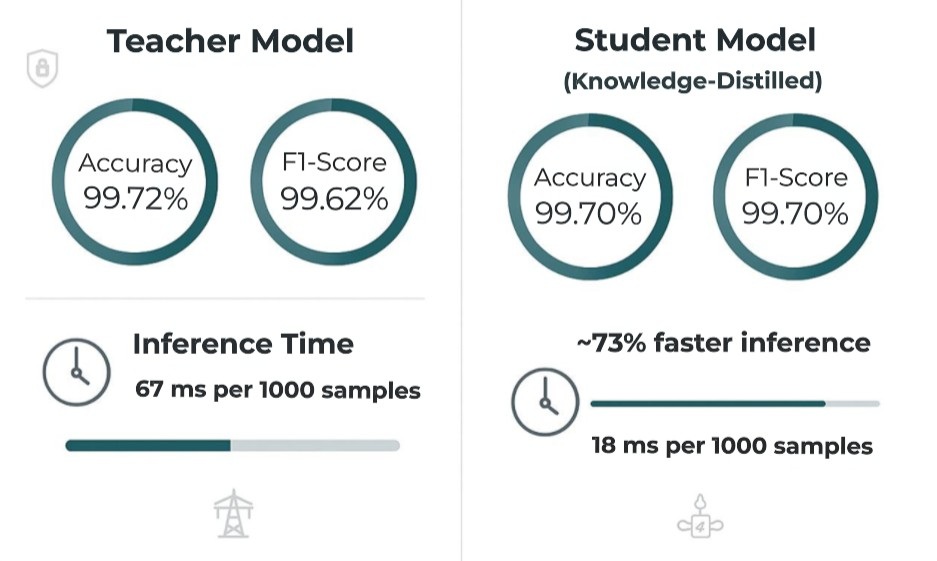}
    \caption{Teacher–student model comparison for microgrid cyberattack detection.}
    \label{fig:Teacher}
\end{figure}

As shown in Fig.~\ref{fig:Teacher}, the student achieved similar results as the teacher model. The knowledge-distilled student preserves diagnostic performance while reducing inference time by approximately 73\% relative to the teacher model. This makes the detector suitable for microgrid edge controllers with limited processing power.

\subsubsection*{Implication}
This demonstrates that knowledge distillation is an effective strategy for deploying cyberattack detection on microgrid controllers with limited computational resources.

\subsubsection*{Time-Domain Prediction Consistency}
\begin{figure*}[!t]
    \centering
    \includegraphics[width=\textwidth]{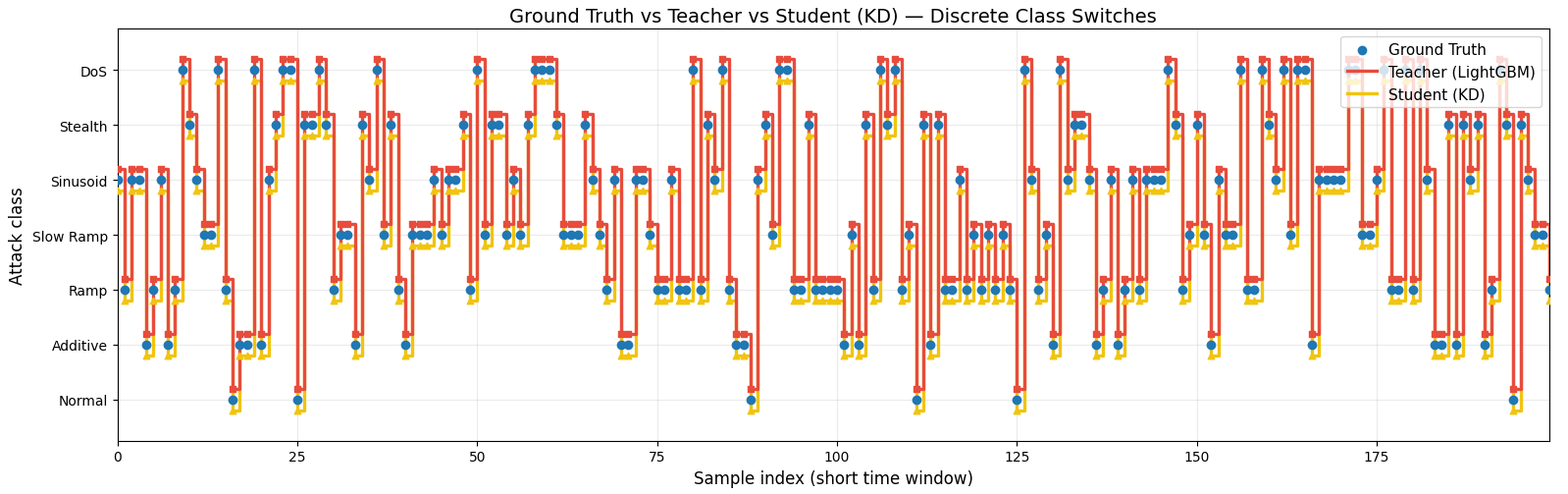}
    \caption{Ground-truth, teacher, and student (KD) discrete class trajectories over a short time window. Blue circles indicate the true class, red squares show the teacher (LightGBM) predictions, and yellow triangles show the student (KD) predictions. A small vertical offset is applied for visual separation. The student closely follows the teacher’s class transitions and matches the ground truth in nearly all samples, indicating that KD preserves temporal decision behavior.}
    \label{fig:kd_switches}
\end{figure*}

In addition to aggregate metrics, we examine whether the distilled student preserves the teacher’s decision behavior over time. Fig.~\ref{fig:kd_switches} plots the ground-truth class sequence and the corresponding teacher and student predictions over a short time-ordered window. A small vertical offset is applied only for visualization, so the three trajectories remain distinguishable. This result supports using the student model for real-time deployment, since it reproduces the teacher’s switching behavior with minimal deviation while providing significantly lower inference time.

\subsection{Real-Time Prediction Behavior}

To assess the real-time decision capability of the proposed intrusion-detection model, a set of ten randomly selected unseen samples was extracted from the held-out test dataset. Because the test set was generated directly from the labeled Simulink scenarios described in Section IV, each sample retains an associated ground-truth attack label that reflects the exact attack mode injected during simulation. These labels were not exposed to the model during training or validation.

\begin{figure}[ht]
    \centering
    \includegraphics[width=\linewidth]{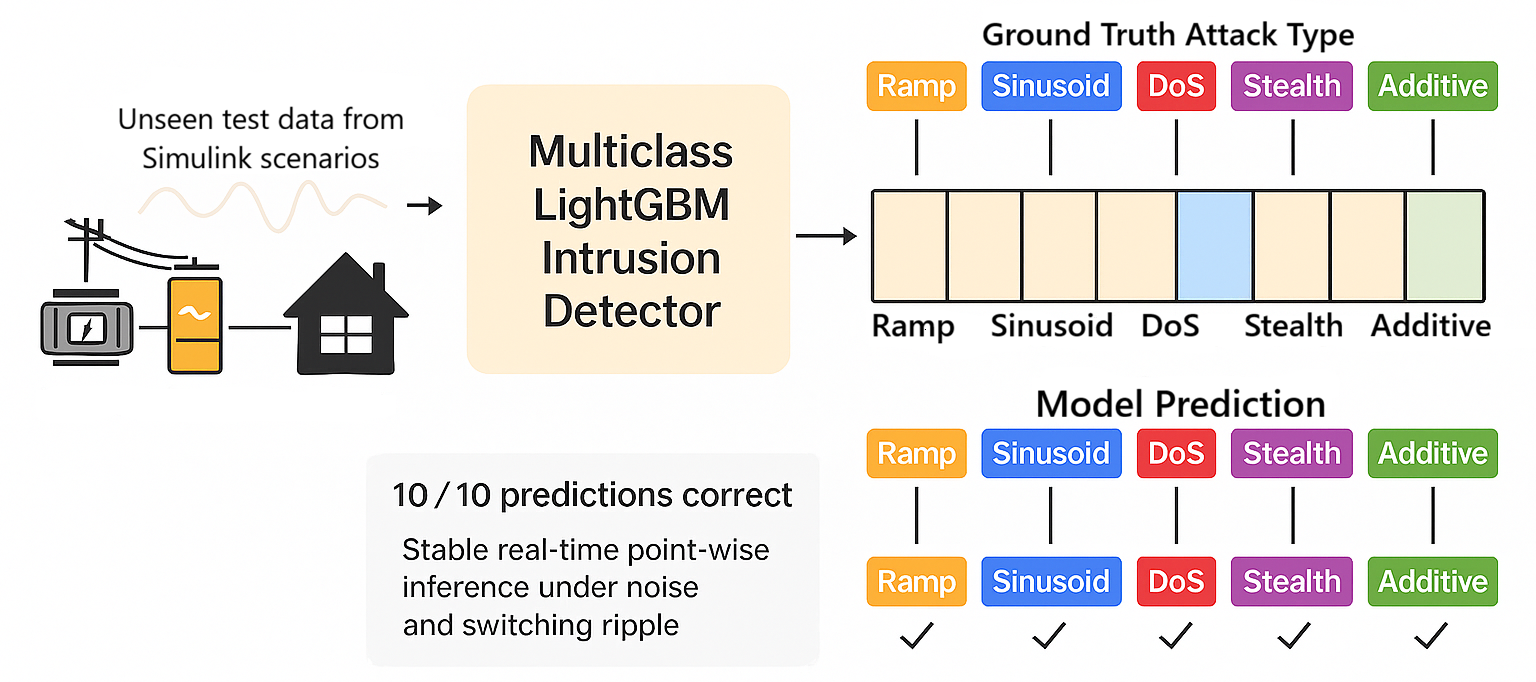}
    \caption{Real-time multiclass attack prediction on unseen test samples.}
    \label{fig:Prediction}
\end{figure}

As shown in Fig.~\ref{fig:Prediction}, the trained multiclass LightGBM model was used to predict each selected sample one at a time. Each prediction was compared with its ground-truth label, and all ten matched correctly. This shows that the classifier works well on unseen data and remains stable under noise and switching ripple. The system is reliable for point-wise real-time inference.

\section{Computational Efficiency and Deployment Considerations}

Effective cyberattack detection in microgrids must be both accurate and fast \cite{krishnamurthy2024tracking}. The detection algorithm must also run with low computational cost and meet strict real-time requirements \cite{munawar2025gwo}. This section evaluates the computational efficiency of the proposed LightGBM-based framework, including model size, inference speed, CPU-only performance, and the benefits of knowledge distillation. All runtime measurements use the same experimental setup as in Section~V.

\subsection{Hardware Environment}
All inference-time and model-size measurements were performed on a virtualized Google Colab CPU instance. The system used an Intel Xeon processor (GenuineIntel) running at 2.20~GHz with one physical core and two hardware threads, based on the \texttt{x86\_64} architecture and virtualized using a KVM hypervisor. The processor provided a 55~MB L3 cache and supported AVX, AVX2, and FMA instruction sets. This environment represents a modest CPU-only computing platform, allowing fair comparison with embedded-class processors that lack GPU acceleration.

\subsection{Model Size and Memory Footprint}

All models, including the binary classifier, multiclass classifier, and distilled student model, were exported in the standard LightGBM text format for size evaluation. Because LightGBM represents each tree using simple threshold-based splits rather than dense numerical tensors, its memory footprint is inherently small compared with deep learning models \cite{ke2017lightgbm}.

Key observations include:
\begin{itemize}

\item Binary LightGBM model: smallest size because it learns only a 2-class boundary.

\item Multiclass LightGBM model: 4.83 MB, reflecting the larger number of classes and trees.

\item Distilled student model: 0.60 MB, an 87.7\% reduction, achieved by using fewer leaves and shallower trees while preserving accuracy.
    
\end{itemize}
These models are much smaller than common deep-learning intrusion detectors, like CNN or LSTM models used for grid event detection, which often range from a few megabytes to hundreds of megabytes \cite{berghout2022machine}.

\subsubsection*{Practical implication}
All three LightGBM models fit easily within the memory limits of low-power embedded hardware, including ARM Cortex-M microcontrollers, TI DSP control boards, Industrial IoT processors and Edge devices like Raspberry Pi units. This small memory size allows the intrusion detection system to run directly on microgrid controllers, without needing cloud support or GPU hardware \cite{bourekouche2025tinyml,  alomari2024lightweight, ojo2025microgrids}.

\subsection{Inference-Time Benchmarking}

Inference latency was evaluated using 1000 consecutive unseen samples from the test dataset. Table IV summarizes the measured times. The binary model is fastest because it performs a single 2-class decision, while the multiclass model is slightly slower due to additional tree evaluations. The student model is the fastest overall, confirming the computational advantage of knowledge distillation.

\subsubsection*{Real-Time Feasibility}

Microgrid control loops typically operate with 100–200 ms update intervals for secondary controllers \cite{guerrero2010hierarchical} and 20–50 ms recommended detection window for cyberattack monitoring \cite{mohammadpourfard2022real}. All models meet these real-time constraints comfortably, even on CPU-only hardware, enabling fully embedded deployment without GPUs.

\subsection{3 CPU-Only Deployment}

A major benefit of the proposed framework is its compatibility with CPU-only environments. Unlike CNNs, LSTMs, or transformer-based detectors that need many floating-point operations and often rely on GPU acceleration, LightGBM uses simple operations like threshold comparisons, integer additions, and shallow tree traversal. These operations are very efficient on embedded processors. In contrast, CNN-based smart grid detectors \cite{hasan2019electricity, wang2016deep} use millions of convolution operations, which are slow and computationally heavy. LSTM-based microgrid models need large matrix calculations and store many internal states in memory. The proposed LightGBM+KD framework avoids these heavy computations and memory costs.

\subsection{ Knowledge Distillation for Edge Intelligence}

KD was incorporated to reduce computational load while preserving classification fidelity. A smaller student LightGBM model was trained to mimic a larger teacher model using softened probability targets and temperature scaling. The results from Section V align with findings in recent edge-intelligence and resource-efficient ML studies \cite{cheng2018model, wu2023knowledge} confirming that KD lowers RAM usage, reduces energy consumption, improves inference speed, increases resilience to noisy measurements, and enables real-time deployment in cyber-physical systems. Thus, KD enhances the practicality of the proposed framework for real-world microgrid controllers.

\subsection{Scalability Considerations} 

The architecture exhibits strong scalability due to the properties of tree-based models and the modularity of the virtual microgrid pipeline.

\subsubsection*{Scalability Drivers}

\subsubsection{Inference does not rely on differential equations}
Unlike physics-based estimators, LightGBM classifiers require no real-time numerical integration.

\subsubsection{Tree-based models scale sublinearly}
Adding new DG measurements or new power-quality features only increases the model size modestly \cite{ke2017lightgbm}.

\subsubsection{Distributed deployment}
Identical local detectors can be placed into each DG controller \cite{hahn2013cyber}.

\subsubsection{Extendable taxonomy}
New cyberattack modes can be integrated by retraining with additional labeled datasets.

\subsubsection{Compatibility with federated learning}
Techniques similar to 
\cite{li2021survey} can be applied so that decentralized microgrids improve the detection model without sharing raw data.

\subsection{Implication for Future Systems}

The results demonstrate that the proposed detection framework is not limited to the specific microgrid modeled in this study. Instead, it provides a scalable foundation that can extend to a broad class of modern cyber-physical power systems. Because the LightGBM-based detectors operate with low computational overhead, require no GPU acceleration, and retain high accuracy under switching noise and communication disturbances, the approach is suitable for deployment in several emerging grid architectures \cite{shabad2021anomaly}.

Firstly, multi-DER distribution grids, which increasingly rely on inverter-interfaced resources, can benefit from having lightweight anomaly detectors co-located with local controllers \cite{mohammadpourfard2022real}. Because the model runs well on CPU-only hardware, it can make real-time decisions even at high sampling rates. Furthermore, inverter-based AC and DC microgrids can add the framework at the secondary or tertiary control level without changing existing control hardware \cite{guerrero2010hierarchical}. This makes the model a practical cybersecurity add-on for power-sharing and voltage-restoration control. Thirdly, the method aligns naturally with edge-to-cloud architectures. A local LightGBM model can run at the edge (e.g., on DG controllers or protection relays), while aggregated attack indicators or distilled student models may be shared with supervisory systems or cloud-hosted analytics platforms \cite{tariq2024fog}.

Finally, the framework is compatible with hybrid AC/DC microgrids that require fast decision-making across heterogeneous converter topologies. The combination of high accuracy, low inference latency, and modest computational requirements makes the proposed system a strong candidate for next-generation resilient smart grids. It enables real-time cyberattack detection across a wide range of microgrid configurations and can be readily extended to new distributed energy systems.

\section{Conclusion}

This work presented a complete cyber-physical framework for generating, labeling, and classifying secondary-control cyberattacks in inverter-based microgrids. A high-fidelity virtual microgrid was built in MATLAB/Simulink using distributed droop-controlled DGs, consensus-based secondary controllers, and structured cyberattack injection pathways. The simulation environment recorded detailed physical and cyber behavior at a very high time resolution of $2~\mu\text{s}$. This made it possible to generate realistic and well-labeled datasets that represent both normal operation and different types of cyberattacks. A complete ML pipeline was built to process the generated datasets. After applying incremental normalization and balanced downsampling, we trained binary and multiclass LightGBM models to detect and classify six representative secondary-control attack types. The multiclass model reached 99.7\% accuracy, showing that the attack types can be recognized correctly by the classifier. KD was then used to compress the model, creating a smaller student classifier that kept the same accuracy while reducing the inference time by about 73\%.

The results show that the tree-boosted models provide a strong mix of accuracy, reliability, and fast computation. Their compact structure enables practical real-time deployment across a wide range of microgrid controllers. Furthermore, the structured dataset and modeling approach provide a repeatable blueprint for future cyber-resilient microgrid studies. Future work will extend the framework to include (i) additional DER types and converter topologies, (ii) communication-layer adversaries with time-varying topologies, (iii) Reinforcement Learning-based adaptive controllers, and (iv) federated or privacy-preserving architectures to enable secure multi-microgrid coordination. Overall, this study shows that using high-resolution cyber-physical simulations together with lightGBM models and KD provides a practical and effective way to detect cyberattacks in real time within modern microgrids.

\section*{Acknowledgments}
Special thanks to Professor Suman Rath for his guidance and mentorship during the AI in Power System course, as well as for his continued supervision and support throughout the doctoral research project.

\bibliographystyle{IEEEtran}   % or plain, ieeetr, unsrt, etc.
\bibliography{references}      % name of your .bib file (without .bib)

\vfill

\end{document}